\documentstyle[twoside]{article}

\font\Bbb = msbm10

\def\HH{{\cal H}}
\def\TT{{\cal T}}
\def\KK{{\cal K}}
\def\BB{{\cal B}}
\def\BC{\mbox{\Bbb C}}
\def\BZ{\mbox{\Bbb Z}}
\def\BR{\mbox{\Bbb R}}
\def\id{\mbox{\Bbb I}}
\def\be{\begin{equation}}
\def\ee{\end{equation}}
\def\bea{\begin{eqnarray}}
\def\eea{\end{eqnarray}}
\def\la{\lambda}
\def\La{\Lambda}
\def\al{\alpha}
\def\vth{\vartheta}
\def\w{\omega}
\def\sz{\sqrt{z}}
\renewcommand\Im{\mbox{Im}}
\newcommand{\QED}{\mbox{\rule[-1.5pt]{6pt}{10pt}}}

\begin{document}
\title{About a resolvent formula}
\author{P. Duclos$^a$,  P. \v S\v tov\'\i\v cek$^b$ and O. V\'a\v na$^b$ \\
{}\\
{\small (a) Centre de Physique Th\'eorique, CNRS, 13288 Marseille--Luminy}\\
{\small and PHYMAT, Universit\' e de Toulon et du Var, BP 132, }\\
{\small 83957 La Garde Cedex, France}\\
{\small (b) Department of Mathematics, Faculty of Nuclear Science, CTU, }\\
{\small Trojanova 13, 120 00 Prague, Czech Republic}
}
\date{}
\maketitle
\begin{abstract}
A resolvent formula, originally presented by Karner in his habilitation
\cite{Kar1}, is discussed. First the formula is considered abstractly and
then it is demonstrated on an explicit example -- the so called simplified
Fermi accelerator.
\end{abstract}
\section{Introduction}
In his habilitation \cite{Kar1} Karner demonstrated on two explicit
examples a resolvent formula
adjusted to the case when the underlying Hilbert space was written as a
tensor
product $\HH_1\otimes\HH_2$ and the  self-adjoint operator in question was
of the form $A_1\otimes\id_2+\id_1\otimes A_2$ plus a perturbation
generally mixing the two factors $\HH_1$ and $\HH_2$.
Usually the spectral decompositions of $A_1$ and $A_2$ are well known.
In \cite{Kar1} Karner calls the formula a modified Krein formula. This is
not quite exact as the formula depends in fact only on the
described particular algebraic structure.
Nevertheless it can be combined effectively with the Krein formula.
Furthermore,
some applications to spectral analysis have been proposed in \cite{Kar1}
however this program doesn't seem to be completed entirely. Moreover the
formula itself has to be extracted from the text of the habilitation.

Nevertheless we believe that
the Karner's formula deserves a more detailed treatment. One reason for it
is that the particular form of the considered operator, as mentioned above,
occurs  in various interesting situations.
As a prominent example we may mention Floquet
Hamiltonians introduced to study time dependent quantum systems
\cite{Howland, Yajima}. Thus in our short contribution we start
from revealing
the algebraic structure of the Karner's formula. To this end  we treat it
abstractly and confine ourselves for a while to a finite-dimensional
case for then all the terms occurring in the formula are well defined.
However this is not so obvious in some concrete examples when the
Hilbert space is infinite-dimensional. It seems that
one has to consider each case separately
in order to verify that the formula actually makes good sense.
Here we discuss as an example the so called simplified Fermi accelerator
\cite{Kar2}.

\section{Karner's formula}
\proclaim Proposition.
Suppose that ${\cal T}$ and ${\cal H}$ are two finite-dimensional
vector spaces, and set ${\cal K}:={\cal T}\otimes{\cal H}$. Furthermore,
let $\{Q_k\}_{k=1}^M$, $\{P_j\}_{j=1}^N$ be two complete sets of projectors
on the vector space $\TT$, i.e.,
$Q_kQ_{k^{'}}=\delta_{kk^{'}}Q_k$, $\sum_k Q_k=\id_\TT$,
$P_jP_{j^{'}}=\delta_{jj^{'}}P_j$, $\sum_j P_j=\id_\TT$.
To a set $\{\lambda_k\}_{k=1}^M$ of complex numbers
and to a set $\{H_i\}_{i=0}^N$ of operators on  ${\cal H}$ we relate
the operators
\bea
& D:=\sum_{k=1}^M\lambda_kQ_k, \\
& K_0:=D\otimes\id_\HH+\id_\TT\otimes H_0, \\
& K:=D\otimes\id_\HH+\sum_{j=1}^N P_j\otimes H_j,\\
& \Lambda(z):=
\sum_{j=1}^N\sum_{k=1}^M P_jQ_k
\otimes \Bigl((H_j+\lambda_k-z)^{-1}-(H_0+\lambda_k-z)^{-1}\Bigr).
\eea
Then it holds
\begin{equation}\label{karner}
(K-z)^{-1}=\Bigl((K_0-z)^{-1}+\Lambda(z)\Bigr){\Bigl(\id+
[D\otimes\id_\HH,\,\Lambda(z)]\Bigr)}^{-1} .
\end{equation}

\noindent{\em Remark}.
$(K-z)^{-1}$, $(K_0-z)^{-1}$ and $\Lambda(z)$ are meromorphic functions
with values in the space Lin$({\cal T}\otimes{\cal H})$,
and all of them converge to $0$ as $|z|\rightarrow +\infty$.
Consequently $\det\Bigl(\id+[D\otimes\id,\Lambda(z)]\Bigr)$
is a meromorphic function as well, with a finite number of poles, and
converging to  $1$ as $ |z|\rightarrow +\infty$. This function
has necessarily a finite number of zeroes and the equality (\ref{karner})
makes sense except of a finite number of points $z\in\BC$.

\smallskip
\noindent{\em Proof}.
Let us assume that $z\in\BC$ is chosen so that all terms in (\ref{karner})
are well defined. Multiplying the relation (\ref{karner})
from the right by the expression
$\Bigl(I+[D\otimes I,\Lambda(z)]\Bigr)(K_0-z)$, and
from the left by the expression $(K-z)$ one arrives at
an equivalent identity, namely
\begin{equation}\label{karner2}
K_0-K=\Biggl(\Lambda(z)(D\otimes I)+
\Bigl(\sum_{j=1}^NP_j\otimes (H_j-z)\Bigr)\Lambda(z)\Biggr)(K_0-z)\, .
\end{equation}
Using the equalities
\bea
&& \Lambda(z)(D\otimes\id)=
\sum_j\sum_kP_jQ_k\otimes\lambda_k\Bigl((H_j+\lambda_k-z)^{-1}-
(H_0+\lambda_k-z)^{-1}\Bigr) ,\quad\nonumber  \\
&& \Bigl(\sum_{j=1}^N P_j\otimes (H_j-z)\Bigr)\Lambda(z)
\nonumber\\
&& \qquad\qquad =\sum_j\sum_kP_jQ_k\otimes (H_j-z)
 \Bigl((H_j+\lambda_k-z)^{-1} - (H_0+\lambda_k-z)^{-1}\Bigr) ,\quad
\nonumber
\eea
one can show that the RHS of (\ref{karner2}) equals
\begin{eqnarray}
&& \Biggl(\sum_j\sum_kP_jQ_k\otimes (H_j+\lambda_k-z)
\Bigl((H_j+\lambda_k-z)^{-1}-(H_0+\lambda_k-z)^{-1}\Bigr)\Biggr)
\nonumber\\
&& \qquad\times\, (K_0-z) \nonumber\\
&& =(K_0-z) - \Bigl(\sum_j\sum_kP_jQ_k\otimes (H_j+\lambda_k-z)
(H_0+\lambda_k-z)^{-1}\Bigr)(K_0-z)
\nonumber\\
&& = -\Bigl(\sum_j\sum_kP_jQ_k\otimes(H_j-H_0)
(H_0+\lambda_k-z)^{-1}\Bigr)(K_0-z)
\nonumber\\
&& = -\Bigl(\sum_jP_j\otimes(H_j-H_0)\Bigr)
\Bigl(\sum_kQ_k\otimes(H_0+\lambda_k-z)^{-1}\Bigr)(K_0-z)
\nonumber\\
&& =(K_0-K)\Bigl(\sum_kQ_k\otimes(H_0+\lambda_k-z)\Bigr)^{-1}
(K_0-z)=K_0-K \, , \nonumber
\end{eqnarray}
and this completes the verification.  \QED

\section{Example: simplified Fermi accelerator}
We set $\TT=L^2([\,0,T\,],dt)$, $\HH=L^2([\,0,1\,],dx)$, and so
$\KK=L^2([\,0,T\,]\times [\,0,1\,],dt\,dx)$. We identify $\KK$,
as usual, with $L^2([\,0,T\,],\HH,dt)$. We set further
$D= -i\partial_t$ with periodic boundary conditions, and
$H_0= -\partial_x^{\,2}$ with Neumann boundary conditions.
Both $D$ and $H_0$ are self-adjoint operators with discrete
spectra. The diagonalization (1) of $D$ is now replaced by an infinite
sum, with $\la_k=k\w$, $k\in\BZ$, where $\w:=2\pi/\,T$, and $Q_k$'s
are the orthogonal projectors on the eigen-functions
$\chi_k(t):=T^{-1/2}\,\exp(i\,k\w t)$. Clearly, the operator
$K_0= -i\partial_t\otimes\id -\id\otimes\partial_x^{\,2}$ is
self-adjoint with a pure point spectrum.

Let us now make a small digression and
consider a perturbation $H_g$ of $H_0$ written in the form sense
as $H_g:=H_0+g\,\tau^\ast\tau$ where  $\tau:{\cal H}\to\BC$ is
the trace operator: $\tau u:=u(0)$. Of course, this means nothing
but that, in the case of $H_g$, the boundary condition at the
point $x=0$ reads $(\partial_x f)(0)=g\,f(0)$ while the boundary
condition at the point $x=1$ is still of Neumann type. In fact,
$H_g$ is an entire analytic
family of type B \cite[ch.VII \S4]{Kato} in the variable $g$.

For a later convenience let us also examine the resolvents of
$H_0$ and $H_g$. We set $R_0(z):=(H_0-z)^{-1}$ and
$R_g(z):=(H_g-z)^{-1}$. It is easy to calculate the Green
function corresponding to $H_0$ explicitly, namely
\be
G_0(x,y)= -\frac{\cos(\sz\, x)\cos\big(\sz (y-1)\big)\,\vth(y-x)
+\{x\leftrightarrow y\} }
{\sz\,\sin(\sz)}
\ee
where $\vth(x)$ is the Heaviside step function. Particularly,
\be
\tau R_0(z)\tau^\ast= -\frac{\cot(\sz)}{\sz} \,.
\ee
The two resolvents are related by the equality
$R(z)=R_0(z) - gR(z)\tau^\star\tau R_0(z)$, and so
\be
R_g(z)-R_0(z)=\frac{-g}{1+g\,\tau R_0(z)\tau^\ast}\,
R_0(z)\tau^\ast\tau R_0(z) \,.
\ee
Here $R_0(z)\tau^\ast\tau R_0(z)$ is a rank-one operator
with the norm
\bea
\|R_0(z)\tau^\ast\tau R_0(z)\|^2 &=&
\tau R_0(\bar z)R_0(z)\tau^\ast\,
\|R_0(\bar z)\tau^\ast\tau R_0(z)\| \nonumber\\
&=& (\tau R_0(\bar z)R_0(z)\tau^\ast)\,
(\tau R_0(z)R_0(\bar z)\tau^\ast) \\
&=& \left(\frac{\Im\,\tau R_0(z)\tau^\ast}{\Im\, z}\right)^2 \,.
\nonumber
\eea

Suppose now that $g(t)$ is a
$T$-periodic real function. Following \cite{Kar2}
we call the time-dependent quantum system determined by the
Hamiltonian $H(t)\equiv H_{g(t)}$ a simplified Fermi accelerator.
The corresponding Floquet Hamiltonian
$K:= -i\partial_t +H(t)$
has the same structure as given in (3) provided one replaces
the sum $\sum_j P_j\otimes H_j$  by the direct integral
\be
\int_0^{T\oplus}H(t)\,dt \,.
\ee
This means that the family of projectors $\{P_j\}_j$
is formally substituted by the spectral decomposition of the
multiplication operator $X\in\BB(\TT)$, $(Xf)(t):=t\,f(t)$, which
has, however, an absolutely continuous spectrum.
Proceeding this way one finds that the definition (4) of the operator
$\Lambda(z)$ has to be modified as follows:
\begin{equation}\label{l}
\Bigl(\Lambda(z){\psi}\Bigr)(t):=
\sum_{k=-\infty}^{+\infty} \chi_k(t)
\Bigl(R_{g(t)}(z-k\w)-R_0(z-k\w)\Bigr)
\int_0^T\overline{\chi_k(\tau)}\,{\psi}(\tau)\,d\tau
\end{equation}
where $\psi\in L^2([\,0,T\,],\HH,dt)$ and $z\in\BC\setminus\BR$.

The definition (12) implies that
\be
\Bigl(\Lambda(z)\,\chi_k\otimes f\Bigr)(t) =
\chi_k(t)\Bigl(R_{g(t)}(z-k\w)-R_0(z-k\w)\Bigr)f,\quad
\forall k\in\BZ\mbox{ and } f\in\HH,
\ee
where we have used the convention
$(u\otimes f)(t):=u(t)\,f$, with $u\in\TT$ and $f\in\HH$.
So $\Lambda(z)$ is densely defined. To show that $\La(z)$ is even
bounded we shall treat the RHS of (12) perturbatively assuming that
$z$ is separated from the real axis, i.e., $|\Im\,z|\ge s_0>0$, and
$g\in C^0$, with the supremum norm $\| g\|_\infty$ being sufficiently
small with respect to $s_0$. Relying on the estimates
\be
|\tau R_0(z)\tau^\ast|=\left|\frac{\cot(\sz)}{\sz}\right|\le
\frac{2}{|\Im\,z|}\sqrt{1+\frac{|\Im\,z|}{4}}\le
\frac{2}{s_0}\sqrt{1+\frac{s_0}{4}} =: \al(s_0)
\ee
and
\be
\|R_0(z)\tau^\ast\tau R_0(z)\|\le
\left|\frac{\tau R_0(z)\tau^\ast}{\Im\,z}\right| \le
\frac{\al(s_0)}{s_0}\,,
\ee
we find that if $\| g\|_\infty<1/\al(s_0)$ then
\bea
\Bigl(\Lambda(z){\psi}\Bigr)(t) &=& \sum_{n=0}^\infty
(-1)^{n+1}\,g(t)^{n+1} \sum_{k=-\infty}^{+\infty} \chi_k(t)\,
\big(\tau R_0(z-k\w)\tau^\ast\big)^n \\
&&\qquad\times\: R_0(z-k\w)\tau^\ast\tau R_0(z-k\w)
\int_0^T\overline{\chi_k(\tau)}\,{\psi}(\tau)\,d\tau \nonumber
\eea
and so
\bea
\|\Lambda(z){\psi}\| &\le& \sum_{n=0}^\infty
\|g\|_\infty^{\:n+1} \left(
\sum_{k=-\infty}^{+\infty}
|\tau R_0(z-k\w)\tau^\ast|^{2n}\, \right.\nonumber\\
&& \  \left. \times\,
\|R_0(z-k\w)\tau^\ast\tau R_0(z-k\w)\|^2\,
\Bigl\|\int_0^T
\overline{\chi_k(\tau)}\,{\psi}(\tau)\,d\tau \Bigr\|_\HH^{\:2}
\right)^{1/2}\qquad \\
&\le& \frac{\|g\|_\infty\,\al(s_0)}
{s_0\big(1-\|g\|_\infty\,\al(s_0)\big)}\,
\|\psi\| \,. \nonumber
\eea

Next let us consider the commutator $[\,D\otimes\id,\,\La(z)\,]$.
One deduces from (13) immediately that
\be
\Bigl([\,D\otimes\id,\,\La(z)\,]\,\chi_k\otimes f\Bigr)(t) =
-i\,\chi_k(t)\,
\partial_t\Bigl(R_{g(t)}(z-k\w)-R_0(z-k\w)\Bigr)f \,.
\ee
Assuming that $g\in C^1$ we get a relation similar to (16), namely
\bea
\Bigl([\,D\otimes\id,\,\La(z)\,]{\psi}\Bigr)(t) &=&
-i\, g'(t) \sum_{n=0}^\infty
(-1)^{n+1} (n+1)\,g(t)^n
\sum_{k=-\infty}^{+\infty} \chi_k(t)\, \nonumber\\
&& \ \times\,
\big(\tau R_0(z-k\w)\tau^\ast\big)^n\,
R_0(z-k\w)\tau^\ast\tau R_0(z-k\w) \qquad\\
&& \ \times\,
\int_0^T\overline{\chi_k(\tau)}\,{\psi}(\tau)\,d\tau \,, \nonumber
\eea
and consequently the estimate
\be
\big\|[\,D\otimes\id,\,\La(z)\,]{\psi}\big\| \le
\frac{\|g' \|_\infty\,\al(s_0)}
{s_0\big(1-\|g\|_\infty\,\al(s_0)\big)^2}\,
\|\psi\| \,.
\ee

Since, apart of the problems with the precise definition of the operator
$\La(z)$, the algebraic structure remains the same as in the
finite-dimensional case we conclude that the formula (5) extends,
as it is, also to our example of simplified Fermi accelerator provided
$|\Im\,z|\ge s_0$, $\|g\|_\infty\,\al(s_0)<1$ and
\be
\frac{\|g' \|_\infty\,\al(s_0)}
{s_0\big(1-\|g\|_\infty\,\al(s_0)\big)^2} <1 \,.
\ee

Of course,  from the point of view of spectral analysis, a truly
interesting result would involve the limit $|\Im\,z|\downarrow0$. There
is no doubt that this goal requires a much more subtle analysis than
that based on the elementary estimate (14), and this program goes
well beyond the scope of this short contribution. We hope anyway
to have shown that the formula (5) may be given a good sense.

\vskip 10pt
 \noindent{\bf Acknowledgements.} P.\,S. wishes to acknowledge
gratefully partial support from the grant
202/96/0218 of Czech Grant Agency.


\end{document}